\documentclass[onecolumn,showpacs,preprintnumbers,amssymb]{revtex4}
\usepackage{graphicx}
\usepackage[bookmarks=true,hyperfigures=true]{hyperref}
\usepackage{color}

\def\beq{\begin{equation}}
\def\eeq{\end{equation}}
\def\bea{\begin{eqnarray}}
\def\eea{\end{eqnarray}}
\def\bit{\begin{itemize}}
\def\eit{\end{itemize}}
\def\pa{\partial}

\def\nn{\nonumber}

\def\Ga{\Gamma}
\def\si{\sigma}

\def\ka{\kappa}

\def\la{\lambda}
\def\La{\Lambda}

\def\vp{\varphi}
\def\vpc{{\varphi_{cl}}}

\def\vp{\varphi}

\def\cL{{\cal{L}}}

\def\cO{{\cal O}}

\makeatletter
\def\underbracket{%
  \@ifnextchar [ %
    {\@underbracket}%
    {\@underbracket [\@bracketheight]}}
\def\@underbracket[#1]{%
  \@ifnextchar [ %
    {\@under@bracket[#1]}%
    {\@under@bracket[#1][0.4em]}}
\def\@under@bracket[#1][#2]#3{
  \mathop {%
    \vtop {%
      \m@th \ialign {%
        ##\crcr $\hfil \displaystyle {#3}\hfil $%
       \crcr \noalign %
       {\kern 3\p@ \nointerlineskip }%
        \upbracketfill {#1}{#2}
       \crcr \noalign %
       {\kern 3\p@ }%
     }%
   }%
  }%
  \limits%
}
\def\upbracketfill#1#2{%
  $\m@th \setbox \z@ \hbox {$\braceld$}
  \edef\@bracketheight{\the\ht\z@}\bracketend{#1}{#2}
  \leaders \vrule \@height #1 \@depth \z@ \hfill
  \leaders \vrule \@height #1 \@depth \z@ \hfill%
  \bracketend{#1}{#2}$%
}
\def\bracketend#1#2{\vrule height #2 width #1\relax}

\begin{document}
\title{Softly broken conformal symmetry\\[1mm]
with quantum gravitational corrections}

\author{Krzysztof A. Meissner$^1$, Hermann Nicolai$^2$ and Jan Plefka$^3$}
\affiliation{$^1$Faculty of Physics,
University of Warsaw\\
Pasteura 5, 02-093 Warsaw, Poland\\
$^2$Max-Planck-Institut f\"ur Gravitationsphysik
(Albert-Einstein-Institut)\\
M\"uhlenberg 1, D-14476 Potsdam, Germany\\
$^3$Institut f\"ur Physik und IRIS Adlershof, Humboldt-Universit\"at zu Berlin, 
  Zum Gro{\ss}en Windkanal 6, 12489 Berlin, Germany\\
}

\begin{abstract}
We show that a previously proposed new mechanism to eliminate quadratic divergences
for scalar masses is self-consistently compatible with corrections induced by perturbative 
quantum gravity, provided the theory embeds consistently into a UV completion at
the Planck scale.
\end{abstract}
\preprint{HU-EP-18/35}
\maketitle

\noindent{\bf Introduction.}
Over the years various schemes have been devised to solve the electroweak hierarchy
problem and thereby to explain the stability of the electroweak scale with regard to the 
Planck scale. Among these, the most prominent proposal is based on low energy
($N\!=\!1$) supersymmetry. However, in the light of recent LHC results showing no 
evidence whatsoever of low energy supersymmetry after a decade of taking data, there 
is clearly a need for alternative ideas. Among these, ans\"atze relying on (some variant of) conformal symmetry have recently received a lot of attention, see
\cite{BA,MN0,Lindner,SW,AMS,Lindner1,noCWyy,OOT,WY,Lindner2} and references therein.
In one way or another, all these proposals aim for a `minimalistic' solution of the 
problem, taking seriously the possibility that the Standard Model (SM) may well survive 
modulo some minor modifications all the way to the Planck scale, 
for which there is now accumulating evidence.

Here we follow up on a recent proposal \cite{CLMN} invoking {\em softly broken 
conformal symmetry} (SBCS) which is equivalent to demanding the cancellation 
of quadratic divergences in terms of bare parameters at a distinguished and very 
large scale  $\La$. This scale serves as an effective cutoff: it is a 
basic assumption that at this scale, a proper theory of quantum gravity `takes over' 
so that the cutoff $\La$ gets identified with a physical scale and is never taken to infinity.
For this reason we usually assume that $\La \sim M_{\rm PL}$. With the assumption that
the pure matter theory is renormalizable, the bare couplings are 
identified with the running couplings evaluated at this distinguished scale:
\beq\label{lbare}
\la_i \equiv \la_i^{\rm bare} =   \la_i(\mu) \Big|_{\mu = \La}
\eeq
The key requirement then reads
\beq\label{SBCS1}
f_i^{\rm quad}(\{\la_j\}) = 0
\eeq
where $f_i^{\rm quad}$ are the coefficient functions of the quadratic 
divergences accompanying the mass renormalizations of the scalar fields, with one such function for each
independent physical scalar field,
\beq\label{deltam}
\delta m_i^2 \,=\,  \left( \frac{\La^2}{16\pi^2} \right)  f_i^{\rm quad} (\{\la_j\})
\,+ \, \cO\big(\log (\La)\big)
\eeq
and where $\{ \la_i \}$ is the collection of all couplings of the model under consideration.
We note that for the unmodified SM the possible vanishing of the quadratic mass divergence
as  a function of the running scalar self-coupling was already investigated in \cite{HKO}, 
with the result that $\La \sim 10^{24}$ GeV, several orders of magnitude above the Planck scale. 
More recently, the above criterion was applied in \cite{CLMN,LMN} to a slightly extended 
version of the SM with right-chiral neutrinos and one extra complex scalar field, 
requiring $\La \sim M_{\rm PL}$, and in such a way that neither 
Landau poles nor instabilities of the effective potential appear up to that scale, 
thus ensuring the survival of the SM essentially {\em as is} up to that scale. 
The term `SBCS' derives its justification from the fact that, as a consequence of 
requiring the absence of quadratically divergent contributions in (\ref{deltam}) 
the physical masses can be kept consistently small in a perturbative treatment
as their quantum corrections depend at worst logarithmically on the cutoff $\La$ 
(we will comment below on the reformulation of this statement in terms of running masses).
Thus, when viewed
from the cutoff scale $\La$, (the matter part of) the theory looks effectively 
conformally invariant because $m_{i}^{2} \lll \Lambda^{2}$.
Indeed, our criterion is somewhat similar to softly broken supersymmetry, 
where one also allows for explicit symmetry breaking terms, on condition that 
these terms do not spoil the cancellation of quadratic divergences.
Our procedure also shows how matter coupled Einstein gravity can give rise to 
a conformally invariant low energy flat space limit (with the gravitational
coupling $\ka\rightarrow 0$) even though Einstein gravity itself is not 
conformally invariant (see also \cite{MN1}).

In this note we wish to show that the above criterion can be maintained in a 
self-consistent manner also if perturbative quantum gravitational corrections are taken 
into account, provided we assume that at the Planck scale the SM (or rather,
some mildly  amended version thereof) merges into a UV complete extension 
(see \cite{QGrefs1,QGrefs2,Don1,Don2,Loebbert:2015eea} and references therein for
previous work concerning perturbative quantum gravity corrections to SM processes).
The main point is that with the assumption of hierarchically
small masses the gravitational corrections to the coefficient functions
$f_i^{\rm quad}$ are of order $(\ka m)^2$, that is
\beq
f^{\rm quad}_i(\{\lambda_{j}\},\kappa)  = \,        
f^{\rm quad}_i(\{\la_j\}) \,+\, \cO\big( (\ka m)^2 \big)  \; ,
\eeq
hence hierarchically smaller than the contribution of the matter couplings (here assumed 
to be $< \cO(1)$, as is the case for all extended SM couplings up to $\cO(M_{\rm PL})$)
over the whole range of energies up to the Planck scale, and thus completely negligible.
The assumed existence of a UV completion is necessary, because otherwise
the theory will be overwhelmed by power law divergences involving arbitrarily high 
powers of $\ka\La$ which, if present, would effectively render moot the whole issue 
of quadratic divergences, as one would expect to be the case for a non-renormalizable 
theory. We refer readers to \cite{SMILGA} for a detailed discussion of the 
adverse effects of such divergences.

Let us also emphasize that, in contrast to Veltman's original proposal \cite{Veltman}, 
our condition of canceling the quadratic divergences with fixed and finite $\La$ is an 
{\em RG invariant statement} in the effective field theory; consequently, in terms of running
couplings, the condition (\ref{SBCS1}) itself becomes scale dependent (and can be
easily obtained by expressing the bare couplings in terms of running couplings \cite{CLMN}).
The whole scheme becomes well defined by the assumed finiteness of the 
Planck scale theory. As far as sub-Planckian physics is concerned, our scheme
also bears some similarity with ideas proposed in the framework of the asymptotic 
safety program (see {\em e.g.} \cite{SW,AMS,Eichhorn,AsymptSafety,GKM} and references therein) 
where the UV completeness would follow from the assumed existence of a  non-trivial 
UV fixed point.

\vskip0.2cm

\noindent {\bf The model: gravity coupled to a set of scalar fields.}
We consider the Lagrangian
\beq
\cL = \frac2{\ka^2} \sqrt{-g} R \,+\, \sqrt{-g} \left( \frac12 g^{\mu\nu} \partial_\mu S^A \partial_\nu S^A 
            - V(S) \right)
\eeq
where $S^A$ are real scalar fields ($A,B,...=1,...,n$) with potential $V(S)$.
We can ignore fermions at this stage, because at the one-loop order considered here
their contribution to $f^{\rm quad}_i$, opposite in sign to the contribution of
the bosonic matter fields, remains the same as in the absence of gravitational 
couplings. By contrast, for scalar fields there appear terms mixing them with $h$, 
and these are ones we have to worry about, see below. For the perturbative expansion 
we split the fields into their background values and fluctuations
\bea 
g_{\mu\nu}(x) &=& \eta_{\mu\nu} + \ka h_{\mu\nu}(x)\;\; ,\quad
h_{\mu\nu} = H_{\mu\nu} + \frac14 \eta_{\mu\nu} h \quad
(\eta^{\mu\nu} H_{\mu\nu} = 0\,,\, h\equiv h^\mu{}_\mu) \nn\\[1mm]
S^A(x) &=& \vp^A_{cl} (x)   + s^A(x)
\eea
with the Minkowski metric $\eta_{\mu\nu}$ and background scalar fields $\vp^A_{cl}(x)$;
$\ka = M_{\rm PL}^{-1}$ is the gravitational coupling.
For establishing the
 mass renormalization it is in fact sufficient to take a static $\vp_{cl}(x)$. The full
Lagrangian also includes a gauge fixing term 
\beq\label{gfterm}
\cL_{gf} = \frac{1}{\xi} \left(\pa^\nu h_{\mu\nu} - \frac12\pa_\mu h\right)^2
\eeq
We will set $\xi =1$ for simplicity (but note that the coefficient functions will depend on the
gauge choice). We neglect gravitational ghost fields as they do not couple to scalar fields 
at the one-loop order, and thus make no contribution to the coefficient of quadratic divergences.
The full action is thus 
\beq
S= \int d^4 x (\cL + \cL_{gf})
\eeq

Our aim then is to perform a perturbative quantization in order to extract the coefficient
of quadratic divergences from both matter and gravitational perturbations. The
effective action is given by 
\beq
\exp\big( i\Ga(\vp_{cl}) \big) = \int dh ds \exp\left( iS(\vp_{cl} + s, \eta + h) \,-\, 
 i \frac{\partial \Ga (g,\vpc)}{\partial \vpc^A}\bigg|_{g=\eta} s^A  \, -
\, i \frac{\partial \Ga (g,\vpc) }{\partial h_{\mu\nu}}\bigg|_{g=\eta}  h_{\mu\nu}  \right)
\eeq
The subtraction removing the linear terms in the fluctuation fields
effectively eliminates tadpoles (see e.g.~\cite{Schwartz:2013pla}). 
Furthermore, at the relevant order 
we can exploit the well known fact that $\Ga(\vpc) = S(\vpc) + \cO(\hbar)$ 
to replace the derivatives of $\Ga$ by derivatives of $S$. Note that it is not 
necessary to impose the equations of motion  on the background fields $\vpc$.

At quadratic order the fluctuations are
\bea
\mathcal{L}'' &=& - \frac12 h_{\mu\nu} P^{\mu\nu;\rho\si} 
           \left(- \Box + \frac{\kappa^2}2 \left [ \frac 12(\partial_{\mu}\vp_{cl})^{2}-V_0\right]\right) h_{\rho\si} \, -       \nn\\[2mm]
&&   - \, \frac12 \ka\,   h \left(- \partial_{\mu}\vp_{cl}\,\partial^{\mu}+V_{0, A} \right)  s^A -
   \, \frac12 s^A \Big[ \delta_{AB} \Box + V_{0,AB}\Big] s^B 
\eea
where 
\beq
V_0 (\vpc) \equiv V \big|_{S=\vpc} \;\;, \quad
V_{0,A}(\vpc)  \equiv \frac{\pa V}{\pa S^A} \bigg|_{S=\vpc} \;\;, \quad
V_{0,AB} (\vpc) \equiv \frac{\pa^2 V}{\pa S^A \pa S^B} \bigg|_{S=\vpc}
\eeq
and
\beq
P^{\mu\nu;\rho\si} \equiv \frac12  \big( \eta^{\mu\nu}\eta^{\rho\si}  -
          \eta^{\mu\rho} \eta^{\nu\si}  - \eta^{\mu\si} \eta^{\nu\rho} \big) \, .
\eeq 
We also note the simplification
\beq
 - \frac12 h_{\mu\nu} P^{\mu\nu;\rho\si}  \left(- \Box + \frac{\kappa^2}2 \mathcal{L}_0\right)  h_{\rho\si}
    = - \frac18 h \left(- \Box + \frac{\kappa^2}2 \mathcal{L}_0\right)  h +
        \frac12  H_{\mu\nu}  \left(- \Box + \frac{\kappa^2}2 \mathcal{L}_0\right)  H^{\mu\nu}
\eeq
where $\cL_0 \equiv \frac12 (\pa_\mu \vpc)^2 - V_0$.
Path integration in quadratic fluctuations leads to the functional determinant 
\beq
\mathcal{M}=\det  \left(- \Box + \frac{\kappa^2}2 \mathcal{L}_0\right)^{-\frac92} \,\cdot \,
\det \left( \begin{array}{cc}    - \Box + \frac{\kappa^2}2 \mathcal{L}_0 \; & \frac{\ka}2 (- \partial_{\mu}\vp_{cl}^B\,\partial^{\mu}+V_{0,B})  \\[2mm]
               \frac{\ka}2 (- \partial_{\mu}\vp_{cl}^A\,\partial^{\mu}+V_{0,A})  \; &   \delta_{AB} \Box + V_{0,AB} 
\end{array} \right)^{-\frac12}
\eeq
where the components are split into 9 traceless modes $H_{\mu\nu}$ (giving the first factor), 
the trace $h$ and the $n$ scalar fields $s^A$ (the contribution of ghosts at this
stage would only supply trivial extra factors of $\det(-\Box)$). Specializing to static 
backgrounds $\vp^A_{cl}=\text{const}$ and Fourier transforming ($\Box=-p^{2}$) 
the last determinant is a degree $n+1$ polynomial in $p^{2}$
\beq
(-)^{n}\det \left ( \begin{array}{cc}    p^{2} - \frac{\kappa^2}2 V_0 \; & \frac{\ka}2 V_{0,B} \\[2mm]
                           \frac{\ka}2 V_{0,A} \; &   -\delta_{AB} p^{2} + V_{0,AB} 
\end{array}  \right )
=(p^{2})^{n+1}-(p^{2})^{n} \left (\frac{\kappa^2}2 V_0 + \sum_{A=1}^{n}V_{0,AA} \right ) +
\mathcal{O}((p^{2})^{n-1}) 
=: \prod_{i=1}^{n+1}(p^{2}-M_{i}^{2})
\eeq
from which we learn that $\sum_{i=1}^{n+1}M_{i}^{2}= \frac{\kappa^2}2 V_0 + \sum_{A=1}^{n}V_{0,AA} $. The effective action in cutoff regularization then follows after Wick rotating as
\beq
i\Gamma(\vpc)=\log \mathcal{M} = -\frac i2
\int_0^\La  \frac{d^4 p_{E}}{(2\pi)^4} 
\left ( 9\log (p^{2}_{E}-\frac{\kappa^2}2 V_0) 
+ \sum_{i=1}^{n+1} \log (p^{2}_{E}-M_{i}^{2}) 
\right )
\eeq
where the subscript $E$ indicates that the integral is to be performed in Euclidean signature.
We first subtract out the zero point energy using
\beq
\int_0^\Lambda \frac{d^4p_E}{(2\pi)^4} \, \ln p^2_E = \frac{\Lambda^4}{32\pi^2}\, (\log \Lambda^2 -\frac 12)\, .
\eeq
In the general case this quartic divergence is multiplied by $(n_B - n_F)$, the difference 
in the number of bosonic and fermionic degrees of freedom. As we assume that the 
UV completion will involve supersymmetry in one way or another, this divergence can 
be ignored. For the determination of quadratic divergences the central integral reads
\beq
\int_0^\La  \frac{d^4 p_{E}}{(2\pi)^4} \log \left( 1 - \frac{m^2}{p^2_{E}} \right)
= - \frac1{16\pi^2} \La^2 m^2 - \frac{m^4}{32\pi^2}\left[ \log \left( -\frac{\La^2}{m^2}\right) - \frac12 \right]
    + \cO(\La^{-2}) \, .
\eeq
Hence the quadratically divergent contributions to the effective action emerging from $\log \mathcal{M}$ 
may be extracted to be \footnote{We note that in a general gauge (\ref{gfterm}) on has
$5\kappa^{2} \to (3+2\xi)\kappa^{2}$ in the above.}
\beq \label{GammaQ}
\Gamma^{\text{div}}(\vpc)= \frac{\Lambda^{2}}{32\pi^{2}}\left(
\frac{9}{2}\kappa^{2}V_{0}+\sum_{i=1}^{n+1}M_{i}^{2}  \right ) + \mathcal{O}(\log\Lambda) =
 \frac{\Lambda^{2}}{32\pi^{2}}\left(
5\kappa^{2}V_{0}+\sum_{A=1}^{n}V_{0,AA} 
 \right )  + \mathcal{O}(\log\Lambda) \, .
\eeq
Here we are only interested in  the mass renormalization (\ref{deltam}), so we need only keep
the terms quadratic in the background fields $\vpc$. We also note that the correction term in 
the second factor of (\ref{GammaQ}) gives only a logarithmically divergent contribution, and can therefore be neglected for our purposes of establishing the quadratically divergent contributions.

Concretely for a general potential
\beq\label{20}
V(S) = \frac12 m^2_{AB} S^A S^B + \frac1{4!} \la_{ABCD} S^A S^B S^C S^D
            + \cO(\ka^2)
\eeq
the relevant divergent contributions to the mass matrix read
\beq
\delta m_{AB}^{2}=
- \left(\frac{\La^2}{16\pi^2} \right) 
  \left(\frac52 \ka^2 m^2_{AB}  + \frac12 \sum_{C=1}^n \la_{CCAB}\right)  + \mathcal{O}(\log\Lambda)
\eeq
From this matrix we can extract the coefficient functions $f^{\rm quad}_i$ simply 
by diagonalization. The main point now is that, after the inclusion of fermionic
contributions and with the assumed absence of quadratic 
divergences, the initial condition $m^2_{AB} \lll \La^2$ implies 
$\ka m  \lll \cO(1)$, whence the initial condition remains consistent with
the quantum gravitational corrections. We thus conclude that the conditions on 
$f_i^{\rm quad}$ are effectively the same as in the flat space theory, as the 
gravitational contribution is hierarchically suppressed.

There appears to be no consensus in the literature whether these statements can or cannot 
be consistently rephrased in terms of running masses. Let us recall that running coupling
parameters are merely an auxiliary, though very convenient, device to parametrize
the scaling behavior of $n$-point correlation functions in renormalizable
quantum field theory, but it is arguable whether the very notion of a running coupling
continues to make sense in the context of non-renormalizable theories \cite{Don1}.
Indeed, while there appears to be no unambiguous way to obtain for the running 
masses $m^2(\mu)$ a quadratic dependence on the scale parameter $\mu$ 
in the context of renormalized perturbation theory (because the subtraction of 
a quadratic divergence leaves ambiguous a finite contribution),
such a dependence can arise in a Wilsonian treatment, where one 
integrates out modes with momenta $\mu^2 < p^2 < \La^2$ (this is the point of
view adopted in asymptotic safety scenarios, see \cite{Eichhorn,AsymptSafety,GKM} and references therein). 
In the latter view our condition (\ref{SBCS1}) would eliminate the quadratic 
dependence of the running masses on $\mu$, and thus ensure a logarithmic running 
of $m^2(\mu)$, keeping the so defined running masses consistently small 
over the whole range of energies $\mu \leq \La$, in accord with the SBCS hypothesis.

Let us also point out that the existence of quadratic (and higher power) divergences
in other parts of the effective action follows directly from the above formulas. For instance, 
to pick out the quadratic divergences in the wave function renormalization 
one only needs to expand the above effective action up to quadratic order in the 
derivatives $\pa_\mu \vpc$. Secondly, possible non-renormalizable interactions 
(with or without derivatives) not written explicitly in the formulas above, i.e.~the 
$\cO(\ka^2)$ terms in ({\ref{20}), will likewise 
pick up power law divergences. However, these will not modify our condition (\ref{SBCS1}), 
but instead affect higher order operators, as already pointed out in \cite{Don1,Don2,Loebbert:2015eea}.

\vskip0.2cm

\noindent{\bf Conclusions.} We have shown that the novel mechanism proposed 
in \cite{CLMN} to avoid the hierarchy problem of the effective quantum field theory 
below the Planck scale can be maintained self-consistently in the presence of 
perturbative quantum gravitational corrections. The main advantage of this proposal 
is that, unlike low energy supersymmetry, it can make do without the extra baggage 
of numerous new, and so far unseen, degrees of freedom and the concomitant  
plethora of new couplings (not to mention the fact that $N\!=\!1$ 
matter coupled supergravities are just as non-renormalizable as 
pure gravity, and therefore eventually will also run into the problem of
power law divergences). Finally, as already pointed out in \cite{CLMN}, and
assuming a minimal extension of the SM  along the lines proposed here can be
validated and the corresponding couplings are known, the condition (\ref{SBCS1}) 
can be subjected to experimental tests.

\vspace{0.4cm}
\noindent{\bf Acknowledgments:} K.A.M.~thanks the AEI for hospitality and financial 
support during this work.  J.P. thanks the AEI for hospitality, his
work is supported through funds of Humboldt-University 
Berlin in the framework of the German excellency initiative.
H.N. is grateful to J. Donoghue for correspondence.

\end{document}